\documentclass[conference]{IEEEtran}
\IEEEoverridecommandlockouts
\usepackage{cite}
\usepackage{amsmath,amssymb,amsfonts}
\usepackage{algorithmic}
\usepackage{graphicx}
\usepackage{textcomp}
\usepackage{xcolor}

\usepackage{hyperref}

\newcommand{\idt}{\texttt{idt}~}
\newcommand{\idtmod}{\texttt{idtmod}~}

\newcommand{\oommf}{\texttt{OOMMF}~}
\newcommand{\oommfc}{\texttt{OOMMFC}~}
\newcommand{\verilog}{\texttt{Verilog-A}~}
\newcommand{\python}{\texttt{Python}~}
\newcommand{\spice}{\texttt{SPICE}~}

\newcommand{\spectre}{\texttt{Cadence\textregistered~Spectre}~}

\newcommand{\sLLGS}{\mbox{s-LLGS}\ }

\begin{document}

\title{A Compact Model for Scalable MTJ Simulation}

\author{
\IEEEauthorblockN{Fernando García-Redondo}
\IEEEauthorblockA{\textit{Arm Ltd, Cambridge, UK} \\
fernando.garciaredondo@arm.com}
\and
\IEEEauthorblockN{Pranay Prabhat}
\IEEEauthorblockA{\textit{Arm Ltd, Cambridge, UK} \\
pranay.prabhat@arm.com}
\and
\IEEEauthorblockN{Mudit Bhargava}
\IEEEauthorblockA{\textit{Arm Inc, Austin, USA} \\
mudit.bhargava@arm.com}
\and
\IEEEauthorblockN{Cyrille Dray}
\IEEEauthorblockA{\textit{Arm Ltd, La Paros, France} \\
cyrille.dray@arm.com}
}

\maketitle

\begin{abstract}
This paper presents a physics-based modeling framework for the analysis and transient simulation of circuits containing Spin-Transfer Torque (STT) Magnetic Tunnel Junction (MTJ) devices. The framework provides the tools to analyze the stochastic behavior of MTJs and to generate \verilog compact models for their simulation in large VLSI designs, addressing the need for an industry-ready model accounting for real-world reliability and scalability requirements. Device dynamics are described by the Landau-Lifshitz-Gilbert-Slonczewsky (\sLLGS) stochastic magnetization considering Voltage-Controlled Magnetic Anisotropy (VCMA) and the non-negligible statistical effects caused by thermal noise. Model behavior is validated against the \oommf magnetic simulator and its performance is characterized on a 1-Mb 28 nm Magnetoresistive-RAM (MRAM) memory product.
\end{abstract}

\begin{IEEEkeywords}
	STT-MRAM, MTJ, Compact modeling, s-LLGS
\end{IEEEkeywords}

\section{Introduction}

Recent advances in MTJ devices \cite{Lee2018, Torunbalci2018a, Zhang2020, Boujamaa2020} open the path to the next generation of system architectures, from embedded battery-less edge devices where off-chip Flash will be replaced by embedded MRAM, to future power-efficient caches in HPC systems. STT-MRAM is being actively developed by foundries and integrated into 28 nm generation CMOS Process Design Kits (PDKs) \cite{Boujamaa2020}.

\begin{figure}[!t]
\centering
\includegraphics[width=\columnwidth]{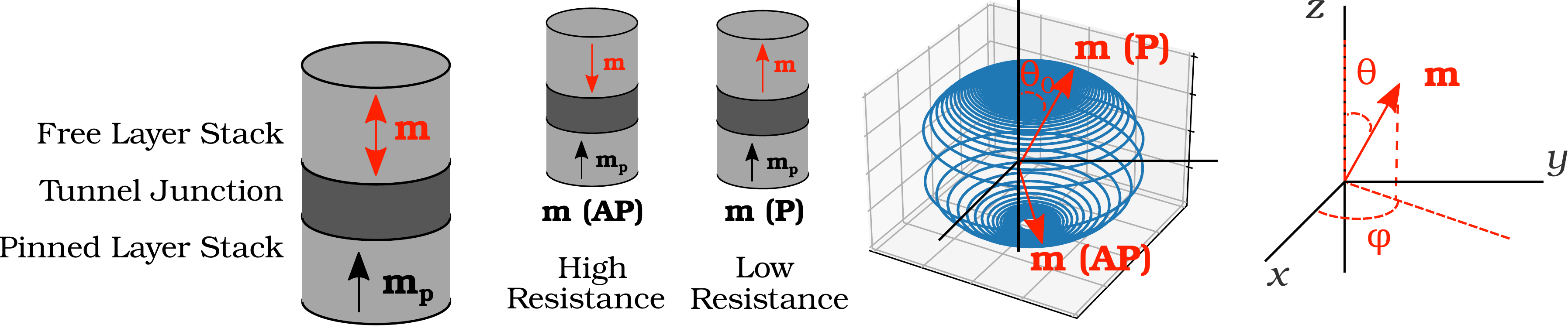}
\caption{Basic MTJ structure, magnetization vector switching trajectory \cite{Ament2016} from Parallel (P) to Anti-Parallel (AP) states, and reference coordinate system.}
\label{fig:mtj}
\end{figure}

The complex multi-layered MTJ structures have heavily non-linear magnetic and electric behaviors that depend on the device structure, the thermal and external magnetic environment, and the applied electrical stress (Figure~\ref{fig:mtj}). On top of this complex system of equations, the evolution of MTJ magnetization $m$ shows a stochastic behavior caused by the random magnetic field induced by thermal fluctuations.

Many MTJ models have been presented, from micro-magnetic \cite{Donahue1999a,Beg2017} approaches to macro-magnetic \spice and \verilog compact models for circuit simulations \cite{Panagopoulos2013, Lee2018, Torunbalci2018a, Zhang2020}. Compact models present in the literature account for different behavioral aspects: a better temperature dependence \cite{Lee2018}, a more accurate anisotropy formulation for particular MTJ structures \cite{Lee2018, Torunbalci2018a, Zhang2020}, and the integration of Spin-Orbit Torque (SOT) for three-terminal devices \cite{Torunbalci2018a, Zhang2020}. These prior works focus on the ability to model a specific new behavior for the simulation of a single or a few MTJ devices in a small circuit. In these approaches, the simulation of device stochasticity imposes a high computational load on the simulation, complicating the simulation of larger VLSI circuits. To address the need to simulate thousands of MTJ devices in a single memory macro, circuit designers may have to oversimplify MTJ switching to a basic interpolated behaviour, and design for worst-case fixed voltage and current targets. This hinders the design of optimized circuits.

Therefore, there is a clear need for an accurate and modular compact model accounting for the magnetization dynamics of MTJ devices yet scalable enough for the simulation of large MRAM memory macros. In this paper we present three key contributions: Section~\ref{sec:motivation} presents a study of reliability issues in MTJ model simulation. The proposed modeling framework is presented in Section~\ref{sec:framework}, including a modular multi-threaded \sLLGS solver validated against \oommf \cite{Donahue1999a}, an efficient \verilog compact model for MRAM macro simulation and a solution to model thermal noise effects. Finally, in Section~\ref{sec:instance} we present a case study with a $1$-Mb $28$ $nm$ MRAM macro.

\section{MTJ Computational and Convergence Issues}
\label{sec:motivation}
From early behavioral prototypes to mature, product-ready PDKs, the simulation of emerging devices is a process that requires not only research contributions capturing novel device behavior \cite{Panagopoulos2013, Lee2018, Torunbalci2018a, Zhang2020}, but also optimization stages enabling circuit design \cite{Garcia2014}. For STT-MRAM, circuit design needs the complex device dynamics to be incorporated into the standard \spice and \spectre \cite{Cadence} solvers. In this section we address the challenges of reliable and efficient simulation of MTJ devices. All results are generated using \texttt{Spectre} $17.1.0.583$ with default settings unless noted otherwise.

The \sLLGS system resolution can easily lead to artificial oscillations if solved in a Cartesian reference system~\cite{Lee2018} when using the Euler, Gear and Trapezoidal methods provided by circuit solvers. In \cite{Ament2016} it is studied how MTJ devices benefit from resolution in spherical coordinates, specially when using implicit midpoint and explicit RK45 methods. Additionally, Cartesian methods require \cite{Lee2018} a non-physical normalization stage for the magnetization vector \cite{Ament2016}. Some further issues arising from solving in a spherical reference system are detailed below.

\begin{figure}[!t]
\centering
\includegraphics[width=\columnwidth]{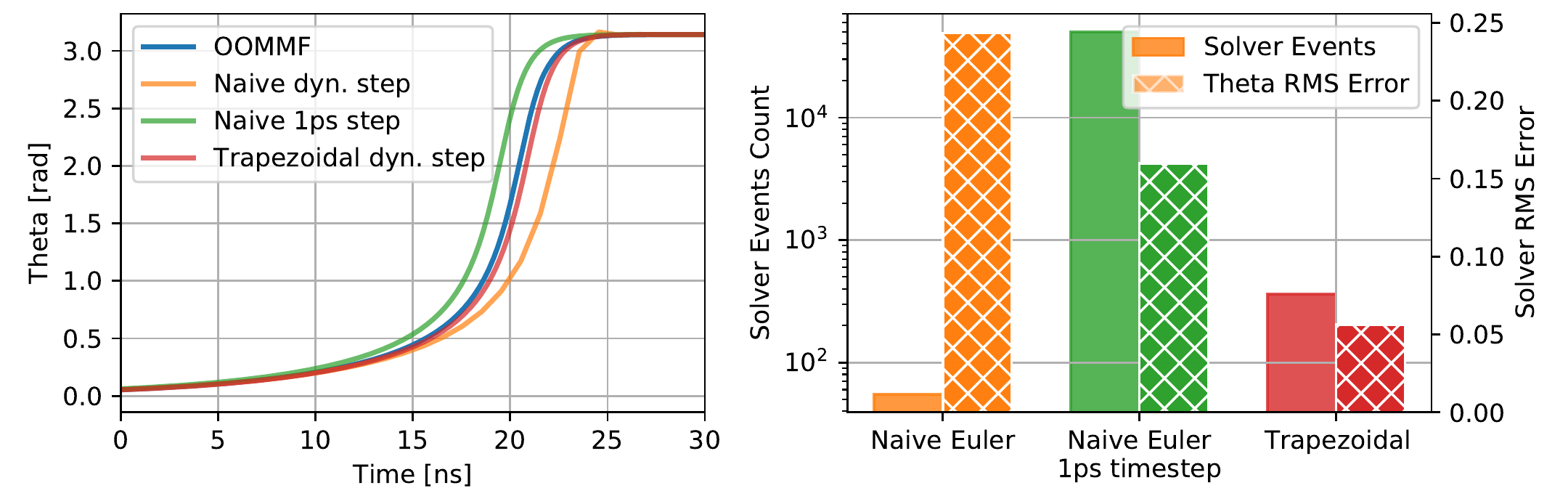}
\caption{MTJ simulation comparing a naive Euler integration against \texttt{Spectre} trapezoidal solver. The naive method leads to larger RMSE (vs. the \oommf reference) even under $1 ps$ timestep with $100\times$ computational load.}
\label{fig:solver_accuracy_dependence}
\end{figure}

\subsection{$\boldsymbol{m_{\phi}}$ Wrapping}
Prior work on MTJ \verilog models uses a naive Euler method to integrate the magnetization vector \cite{Lee2018, Zhang2020}
\begin{equation}
	\boldsymbol{m}(t) = \boldsymbol{m}(t-dt) + \boldsymbol{dm}dt.
	\label{eq:naive}
\end{equation}
However, Equation~\ref{eq:naive} causes $\boldsymbol{m_{\phi}}$ to exceed the $[0, 2\pi)$ physically allowed range, and grow indefinitely over time, leading to unnatural voltages that eventually prevent simulator convergence. Our solution is to use the circular integration \texttt{idtmod} function provided by \texttt{Spectre} \cite{Garcia2014, Cadence} which wraps $\boldsymbol{m_{\phi}}$ at $2\pi$ and prevents it from increasing indefinitely.

\subsection{$m_{\theta}$ Accuracy and Minimum Timestep}
Vector $\boldsymbol{m_{\theta}}$ represents the binary state of the MTJ data bit and is the most critical component to accurately model MTJ switching. The naive integration of $\boldsymbol{m_\theta}$ directly encoding Equation~\ref{eq:naive} in the \verilog description incurs a substantial error during a switching event, as shown in Figure~\ref{fig:solver_accuracy_dependence}. Reducing the timestep improves the RMS error but at the cost of a large increase in computational load. Our solution is to use the \texttt{idt} function for the circuit solvers provided by \texttt{Spectre} \cite{Cadence}, which adapt the integration based on multiple evaluations of the differential terms at different timesteps, leading to better accuracy. Figure~\ref{fig:solver_accuracy_dependence} describes the accuracy RMSE when using a naive Euler method and the \texttt{idt} trapezoidal solver, when compared against \oommf reference. The graph highlights how even with a smaller timestep involving $100\times$ more computed events, the naive integration leads to larger errors.

\begin{figure}[!t]
\centering
\includegraphics[width=\columnwidth]{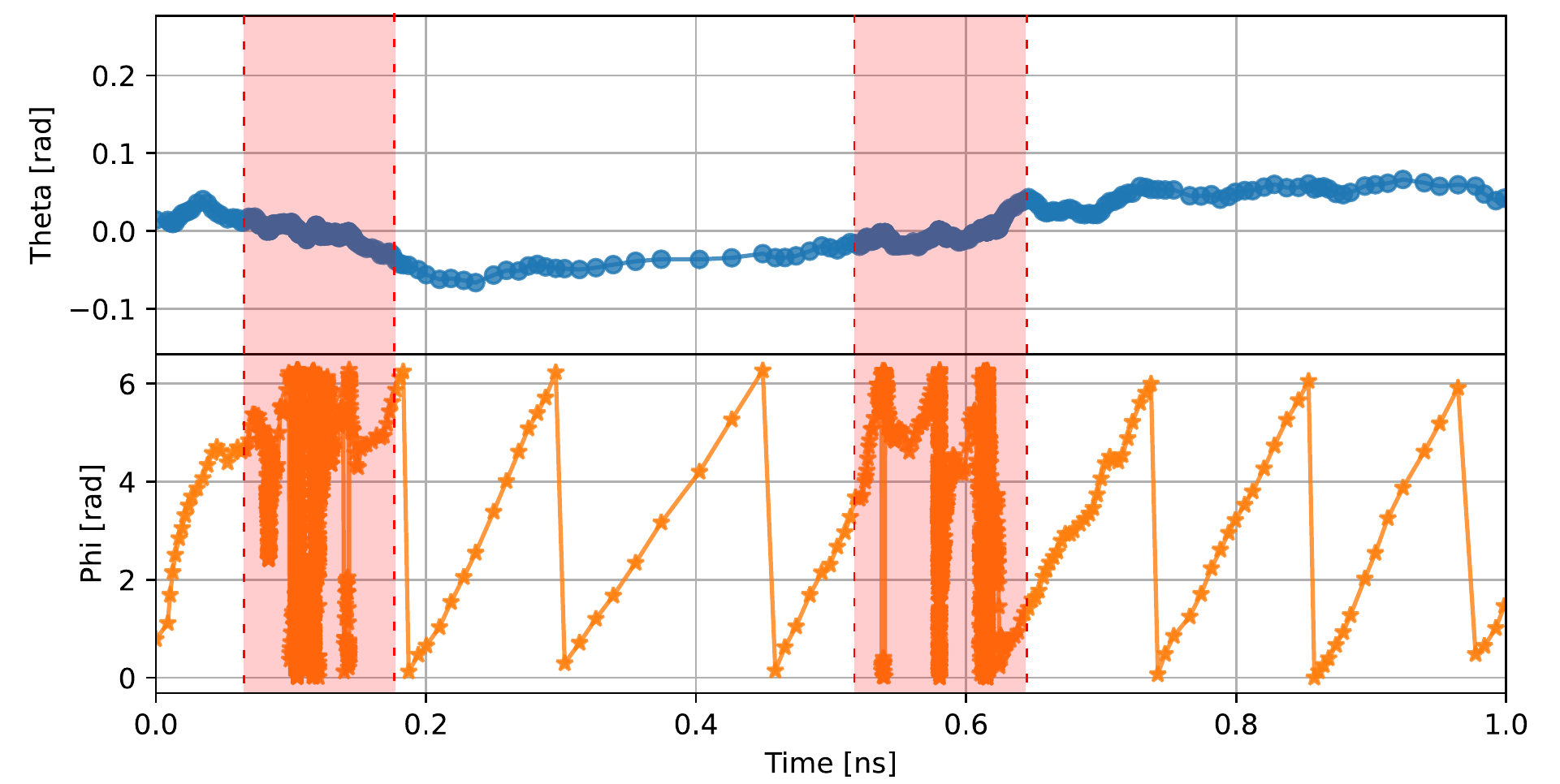}
\caption{
	MTJ \sLLGS simulation under no external current/field, considering thermal $\boldsymbol{H_{th}}$ effects. The evolution of $\boldsymbol{m_{\phi}}$ and $\boldsymbol{m_{\theta}}$ over time highlights the asymptote on $\frac{d}{dt}\boldsymbol{m_\phi}$ at $\theta=0$. The trapezoidal solver under relaxed tolerance successfully provides the desired solution while using a dynamic time step.
}
\label{fig:mtj_phi_acceleration}
\end{figure}

\subsection{$m_\phi$ acceleration}
If a larger circuit is to be simulated and fixed timesteps banned, to avoid instabilities it is essential to manage the tolerance/timestep scheme \cite{Ament2016}, especially during the computation of the $\boldsymbol{m_{\phi}}$ component. The asymptote $\frac{d}{dt}\boldsymbol{m_\phi} \to \infty$ at $\boldsymbol{m_\theta}=0$ accelerates the precession mechanism \cite{Ament2016}. As described by Figure~\ref{fig:mtj_phi_acceleration}, this requires the solver to accordingly increase its resolution. A bounded or fixed time step would fail either to provide sufficient resolution or sufficient performance.

\section{Proposed Framework}
\label{sec:framework}
\begin{figure}[!t]
\centering
\includegraphics[width=\columnwidth]{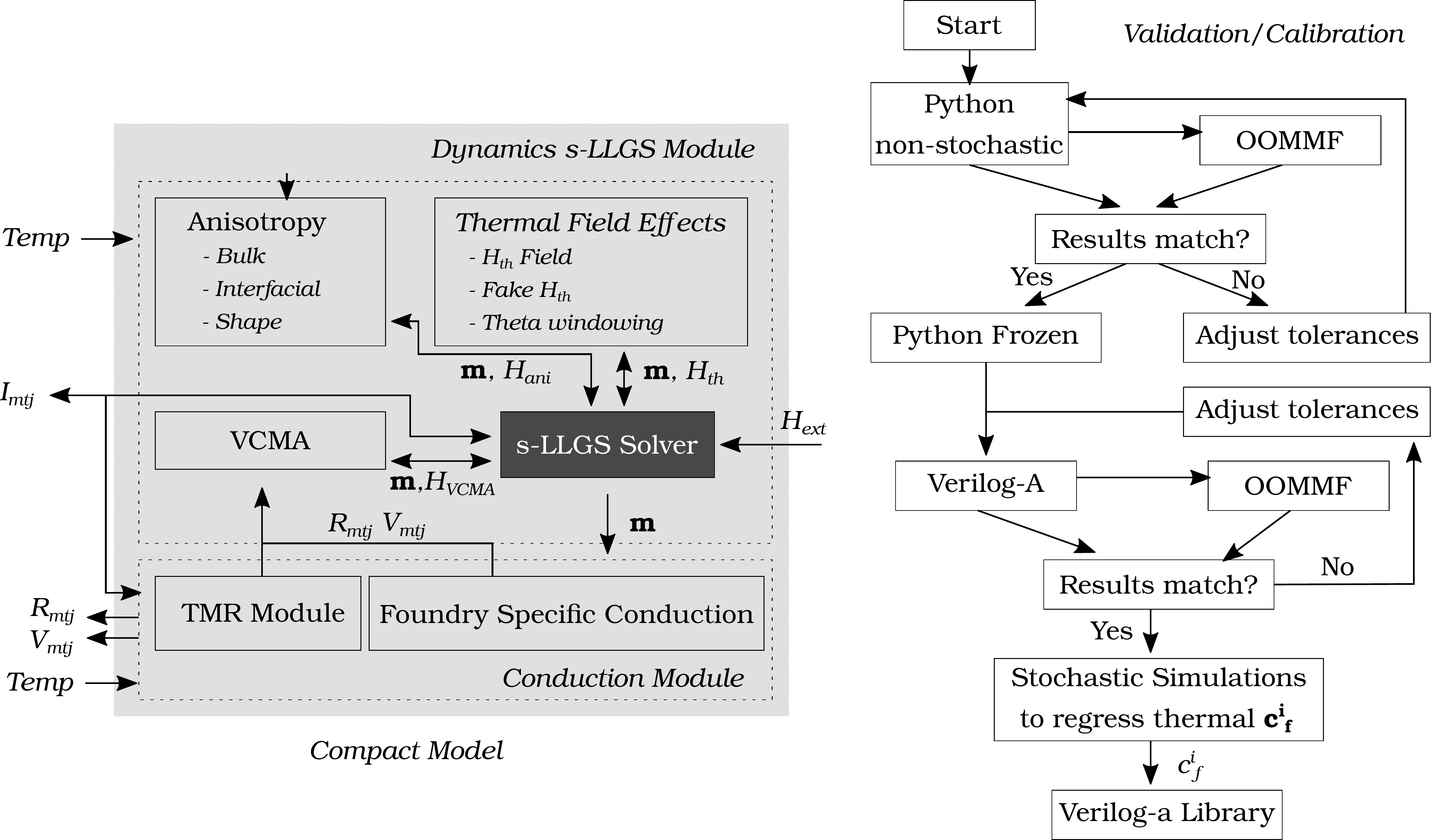}
\caption{
Proposed MTJ modeling framework. The \emph{Conduction} and \emph{Dynamics modules} determine the device conduction and the magnetization based on the external, device anisotropy, STT and thermal induced fields. The \python and \verilog models validate the MTJ behavior against \oommfc.
}
\label{fig:solution}
\end{figure}

In this section we describe the implementation and validation of the proposed MTJ compact model addressing the challenges described in Section~\ref{sec:motivation}, emphasizing its \emph{dynamics module} computing the MTJ magnetization and proposing solutions to incorporate thermal noise effects. 

\subsection{Compact Model Structure}
\label{sec:compact_model}
Figure~\ref{fig:solution} describes the implemented compact model, composed of two modules: the \emph{Conduction} and \emph{Dynamics} modules. The conduction scheme describing the instantaneous MTJ resistance is dependent on the foundry engineered stack. Our modular approach allows foundry-specific conduction mechanisms to complement the basic Tunnel-Magneto-Resistance (TMR) scheme \cite{Graduate2020,Lee2018,Zhang2020,Torunbalci2018a}. The \emph{Dynamics} module describes the temporal evolution of the MTJ magnetization $\boldsymbol{m}$ as a monodomain nanomagnet influenced by external and anisotropy fields, thermal noise and STT \cite{Donahue1999a, Ament2016}, described by the \sLLGS equations \cite{Donahue1999a} in I.S.U.
\begin{eqnarray}
	\frac{d \boldsymbol{m}}{dt} = & -\gamma' \boldsymbol m \times \boldsymbol{H_{eff}} \nonumber + \alpha \gamma' \boldsymbol{m} \times \frac{d \boldsymbol{m}}{dt} \nonumber \\
		& + \gamma' \beta \epsilon ( \boldsymbol{m} \times \boldsymbol{m_p} \times \boldsymbol{m}) - \gamma' \beta \epsilon' (\boldsymbol{m} \times \boldsymbol{m_p}) \nonumber
\end{eqnarray}
\begin{eqnarray}
	\beta = |\frac{\hbar}{\mu_0 e}| \frac{I}{V M_s}, \epsilon = \frac{P \Lambda^2}{(\Lambda^2+1) + (\Lambda^2-1)(\boldsymbol{m} \cdot \boldsymbol{m_p})}
\end{eqnarray}
where $alpha$ and $\gamma$ are the Gilbert damping factor and gyromagnetic ratio respectively, related by $\gamma' = \frac{\gamma \mu_0}{1+\alpha^2}$, $P$ is the polarization factor, $M_s$ is the magnetization saturation, $I$ is the current flowing through the MTJ volume $V$ , $m_p$ is the pinned-layer unitary polarization direction and $\Lambda$ and $\epsilon'$ set the primary and secondary spin transfer terms respectively. The effective magnetic field for a Perpendicular Magnetic Anisotropy (PMA) is defined by the anisotropy field, the external field and the thermal induced field $\boldsymbol{H_{eff}} = \boldsymbol{H_{ani}} + \boldsymbol{H_{ext}} + \boldsymbol{H_{th}}$. The anisotropy field is composed of the PMA uniaxial term, the shape anisotropy demagnetization field, and the VCMA field \cite{Lee2018, Zhang2020}, becoming the $\boldsymbol{H_{eff}}$ vector
\begin{eqnarray}
	\boldsymbol{H_{eff}} = & \boldsymbol{H_{ext}} + \boldsymbol{H_{uni}} + \boldsymbol{H_{demag}} - \boldsymbol{H_{VCMA}} + \boldsymbol{H_{th}}  \nonumber \\
	= & \boldsymbol{H_{ext}} + \frac{2 K_i}{t_{fl} \mu_0 M_s} \boldsymbol{m_z} - M_s \boldsymbol{N} \cdot \boldsymbol{m} \nonumber \\
		  & - \frac{2 \xi I R_{mtj}}{t_{fl} t_{ox} \mu_0 M_s} m_z \boldsymbol{z} + \boldsymbol{\mathcal{N}(0, 1)} \sqrt{\frac{2 K_B T \alpha}{\gamma' M_s V \Delta_t}},
	\label{eq:sllgs}
\end{eqnarray}
where $K_i$ is the interfacial energy constant, $t_{fl}$ and $t_{ox}$ are the free layer and oxide thicknesses, $\boldsymbol{N}$ is the shape anisotropy demagnetization factor, $\xi$ the VCMA coefficient, $K_B$ the Boltzmann constant and $\boldsymbol{\mathcal{N}(0, 1)}$ a Gaussian random vector with components in $\boldsymbol{x, y, z}$ meeting \cite{Borjas2010, Ament2016} conditions.

The compact model has been implemented in \python by adapting the \texttt{Scipy solve\_ivp} \cite{Virtanen2020} solvers to support the $\boldsymbol{H_{th}}$ simulation as a pure Wiener process. The parallel \python engine enables MC and statistical studies. 
The \verilog implementation uses \idt/\idtmod integration schemes with parameterizable integration tolerances.

\subsection{Validation Against OOMMF}
\begin{figure}[!t]
\centering
\includegraphics[width=\columnwidth]{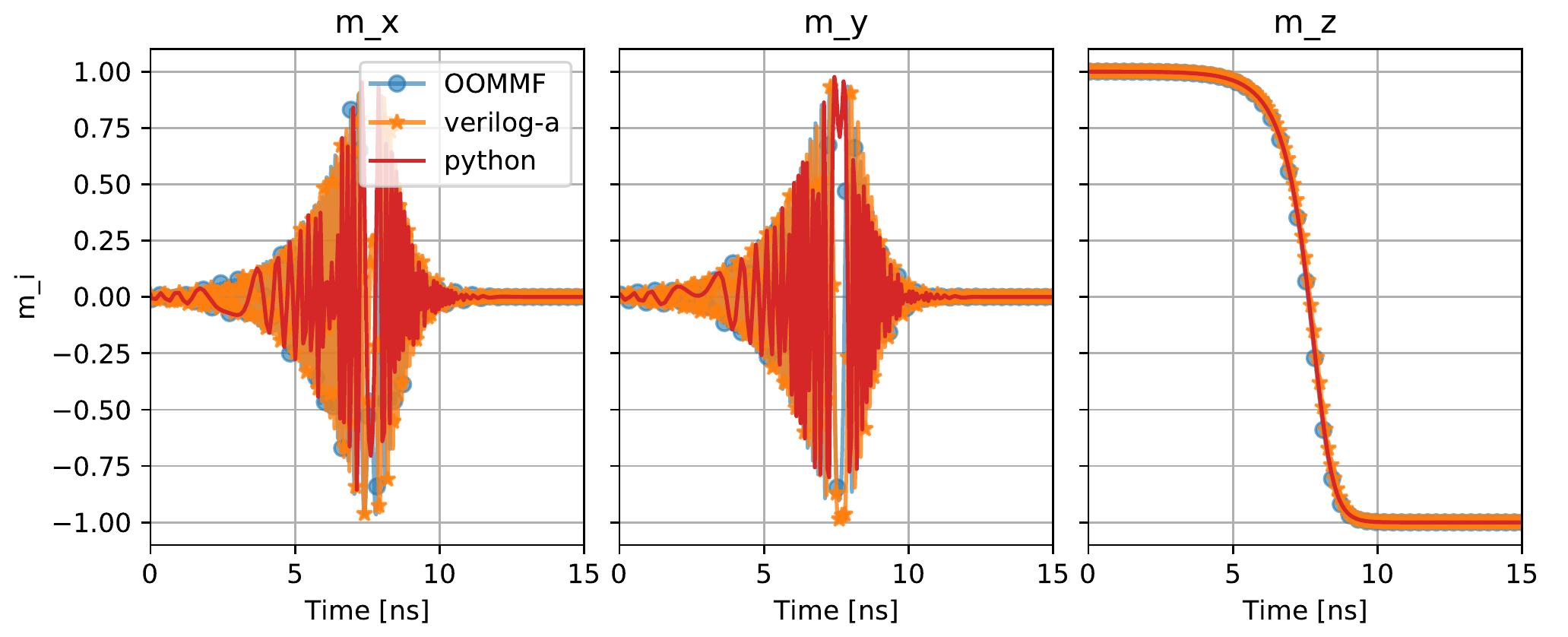}
\caption{
	Validation against OOMMF Framework. A cylindrical MTJ with $50$ $nm$ diameter, $1$ nm thickness, $K_i=1e-3$ $J m^{-2}$, $P=0.75$, $\alpha=0.01$, $\gamma=1.76e11$ $rad/s/T$, $M_s=1.2$ $A m^{-1}$ switches by a current of $35uA$.
}
\label{fig:validation}
\end{figure}

The compact model is validated against the NIST \oommf MicroMagnetic Framework \cite{Donahue1999a}, testing our results against its \oommfc interface \cite{Beg2017}. Following the method from \cite{Ament2016}, we use \texttt{SpinXferEvolve} to simulate a single magnet under an induced spin current. Figure~\ref{fig:validation} shows that \python and \verilog implementations compare well with \oommf with adaptive tolerance driven computations.

\subsection{Thermal noise and MTJ stochasticity}
The random magnetic field $\boldsymbol{H_{th}}$ caused by thermal fluctuations induces stochastic MTJ behaviour resulting in a non-zero Write Error Rate (WER) upon switching events. $\boldsymbol{H_{th}}$ follows a Wiener process \cite{Borjas2010, Ament2016}, in which each $x, y, z$ random component is independent of each other and previous states. This implies that the computation of $H_{th}$ requires large \emph{independent} variations between steps, hindering the solver's attempts to guarantee signal continuity under small tolerances.
The scenario shown in Figure~\ref{fig:mtj_phi_acceleration} leads to computational errors under default solver tolerances,
and excessive computational load under 1 ps bounded time steps.

\begin{figure}[!t]
\centering
\includegraphics[width=\columnwidth]{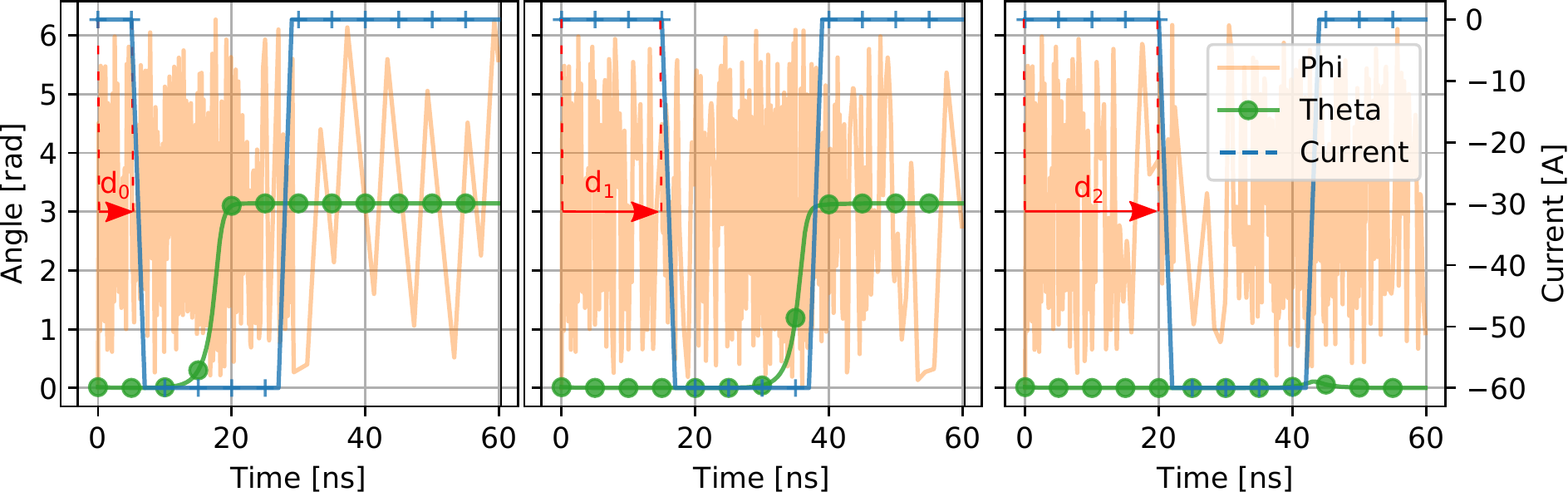}
\caption{
	Damping effect on $\boldsymbol{m_\theta}$ under $\boldsymbol{H_{th}}$ field absence.
	The absence of STT-current (during the $d_2$ delay) collapses $\boldsymbol{m_\theta}$ preventing MTJ switching.
}
\label{fig:mtj_damping}
\end{figure}

Three solutions have been proposed in the literature. First, to emulate the random field by using an external current or resistor-like noise source \cite{Panagopoulos2013}. However, \spice simulators impose a minimum capacitance on these nodes filtering the randomness response, therefore preventing a true Wiener process from being simulated. Second, to bound a fixed small timestep to the solver \cite{Lee2018, Torunbalci2018a}, but as described before this is not feasible for large circuits. Third, to only consider scenarios where the field generated by the writing current is much larger than $\boldsymbol{H_{th}}$, forcing $\boldsymbol{H_{th}}=0$ \cite{Lee2018, Zhang2020}. This has strong implications when moving from single to multiple successive switching event simulations. Under no $\boldsymbol{H_{th}}$ thermal field, the magnetization damping collapses $\boldsymbol{m_\theta}$ to either $0$ or $\pi$. \sLLGS dynamics imply that the smaller the $\boldsymbol{m_\theta}$ the harder it is for the cell to switch, and if completely collapsed, it is impossible. This artificial effect, depicted in Figure~\ref{fig:mtj_damping}, does not have an equivalent in reality, as $\boldsymbol{H_{th}} \neq 0$ imposes a random angle. Design, validation and signoff for large memory blocks with integrated periphery and control circuits requires the simulation of sequences of read and write operations, with each operation showing identical predictable and switching characteristics. However, the damping artifact discussed above prevents or slows down subsequent switches after the first event, since the subsequent events see an initial $\theta$ value vanishingly close to zero. Two solutions are proposed below.

\subsubsection{Windowing Function Approach}
Our first objective is to provide a mechanism for $\boldsymbol{m_\theta}$ to saturate naturally to the equilibrium value given by $\boldsymbol{H_{th}}$ during switching events. By redefining the evolution of $\boldsymbol{m}$ on its $\theta$ component we are able to saturate its angle at $\theta_0$, the second moment of the Maxwell-Boltzmann distribution of $\boldsymbol{m_\theta}$ under no external field \cite{Panagopoulos2013, Lee2018, Torunbalci2018a, Zhang2020}. The new derivative function $\frac{d}{dt}\boldsymbol{m_\theta'}$ uses a Tukey window with the form $\frac{d}{dt}\boldsymbol{m_\theta'} = w_{tukey}(\boldsymbol{m_\theta}, \theta_0) \frac{d}{dt}\boldsymbol{m_\theta}$ where
\begin{equation}
	w(m_\theta) = \left\{
	   \begin{array}{lcc}
		   0 & \mbox{if} & m_\theta < \theta'_0 \\
		   0.5 - \frac{cos}{2}(\frac{4 \pi (m_\theta-\theta'_0)}{\theta'_0}) & \mbox{if} & m_\theta - \theta'_0 < \frac{\theta'_0}{4} \\
	   1  & & \mbox{otherwise}
 	   \end{array}
	   \right.
\end{equation}
defined for $m_\theta \in [0, \frac{\pi}{2}]$ and defined symmetrically ($w'(m_\theta) = w(\pi-m_\theta)$) for $m_\theta \in [\frac{\pi}{2}, \pi]$. This function slows down $\frac{d}{dt}m_\theta'$ when reaching the angle $\theta'_0 = c_w \theta_0$, therefore saturating $m_\theta$ and avoiding $\boldsymbol{m}$ from collapsing over $\boldsymbol{z}$. Moreover, by using $c_{w}^{mean}$ we are able to define the angle that statistically follows the mean stochastic MTJ behavior. Similarly, by simply using the set of parameters $[c_{w}^{worst}, c_w^{best}, c_w^{WER_0}, ..., c_w^{WER_i}]$ we can simulate the worst, best, $WER_i$ behaviors, analyzing how a given circuit instantiating that MTJ device would behave statistically with negligible simulation performance degradation.

\subsubsection{Emulated Magnetic Term}
\label{sec:hfth}
The windowing function approach prevents artificial $\boldsymbol{m_\theta}$ saturation at the end of switching events, but still does not capture the mean effect of $\boldsymbol{H_{th}}$ during switching or under low-current excitation, such as during read events. Our next objective is to address this without performing a large ensemble of random transient simulations. The expansion of Equation~\ref{eq:sllgs} after its expression in spherical coordinates describes $\boldsymbol{m_\theta}$ evolution as proportional to $\boldsymbol{H_{eff \phi}} + \alpha \boldsymbol{H_{eff \theta}}$, leaving $\frac{d}{dt}\boldsymbol{m_\theta} \simeq \frac{\gamma'}{1+\alpha^2} \boldsymbol{H_{eff \phi}}$. We propose to add a fictitious $\boldsymbol{H_{fth} \phi}$ term $\boldsymbol{H_{eff}}$ with the purpose of emulating the mean/best/worst statistical $\boldsymbol{H_{th}}$ contribution that generates $\theta_0$ \cite{Panagopoulos2013, Lee2018, Torunbalci2018a, Zhang2020}. By defining 
\begin{equation}
	\boldsymbol{H_{fth}} = c_f  \sqrt{\frac{2 K_B T \alpha}{\gamma' M_s V \Delta_t}} \boldsymbol{\phi}
\end{equation}
we are able to efficiently model -- by simply using $[c_f^{worst}, c_f^{best}, c_f^{WER_0}, ..., c_f^{WER_i}]$ set of parameters -- the statistical behavior caused by the thermal effects while avoiding their inherent simulation disadvantages.

\begin{figure}[!t]
\centering
\includegraphics[width=\columnwidth]{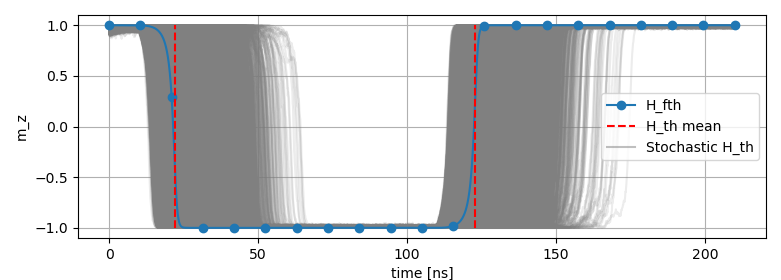}
\caption{
	$10^5$ stochastic simulations for the calibration of $\boldsymbol{H_{fth}}$.
}
\label{fig:calibration}
\end{figure}
\begin{figure}[!t]
\centering
\includegraphics[width=\columnwidth]{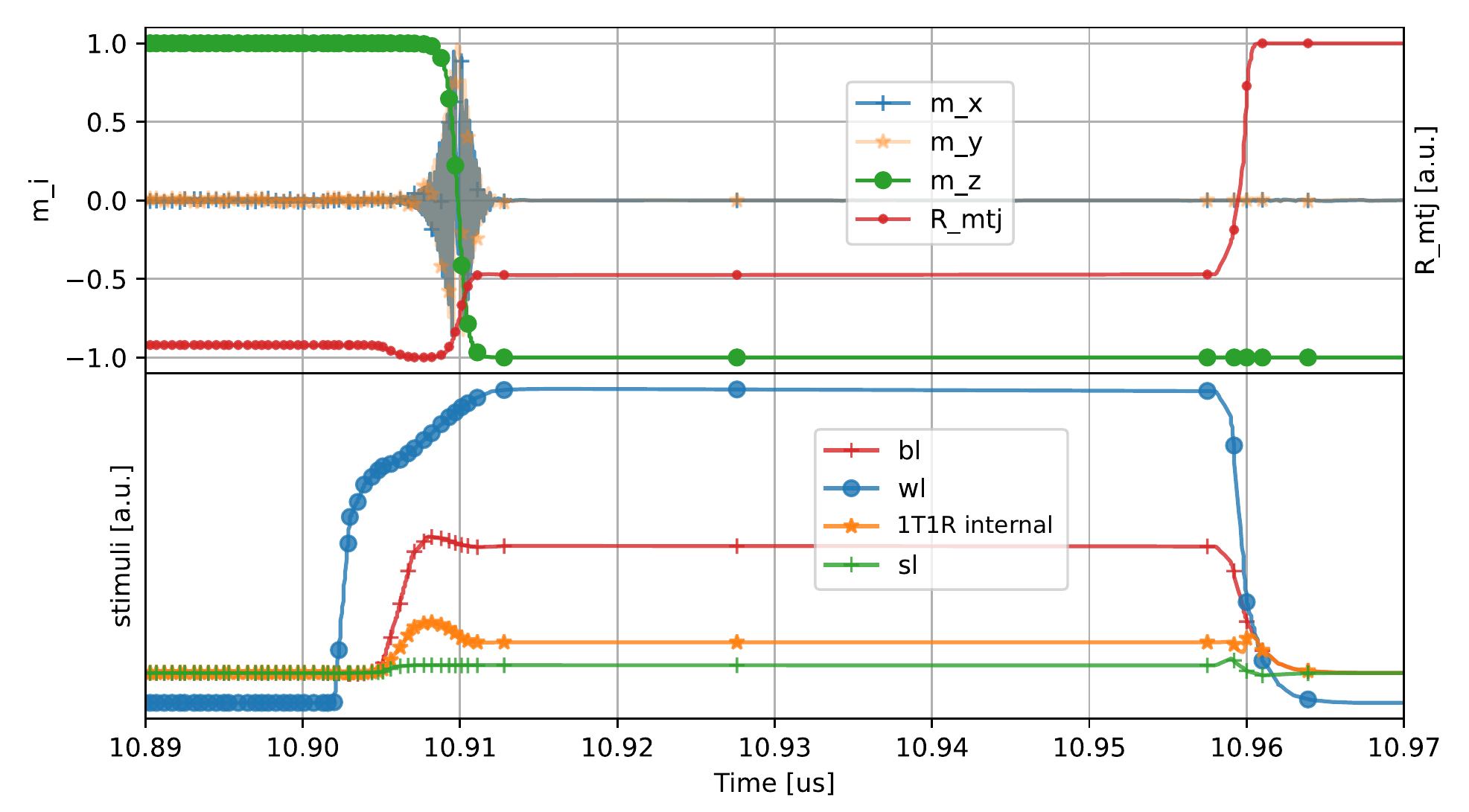}
\caption{Memory macro simulation showing MTJ switching.}
\label{fig:macro}
\end{figure}

The process of calibrating a particular compact model instance is described in Figure~\ref{fig:solution}, and ensures the validation of \python and \verilog instances against \texttt{OOMMF}, before the regression of $c_f^i$ or $c_w^i$ coefficients takes place with a one-off stochastic simulation step as shown in Figure~\ref{fig:calibration}.


\section{$1$-Mb MRAM Macro Benchmark}
\label{sec:instance}
To validate scalability on a commercial product, the model is instantiated into the the $64\times4$ memory top block
of the extracted netlist from a $1$-Mb $28$ $nm$ MRAM macro \cite{Boujamaa2020}, and simulated with macro-specific tolerance settings. The emulated magnetic term from Section~\ref{sec:hfth} enables the previously impossible capability of simulating successive writes with identical transition times due to non-desired over-damping.
Figure~\ref{fig:macro} --resistance/voltage units omitted for confidentiality-- describes a writing operation $10$ $\mu s$ after power-on sequence.
We combine the \sLLGS \oommf validated dynamics with foundry-given thermal/voltage conductance dependence,
providing the accurate resistance response over time.
Compared to using fixed resistors, there is an overhead of $3.1 \times$ CPU time and $1.5\times$ RAM usage. In return, circuit designers can observe accurate transient switching behaviour and read disturbs.

\section{Conclusions}
This work presents an MTJ modeling approach for large VLSI circuits. We analyze MTJ modeling challenges and propose solutions to accurately capture stochastic thermal noise effects. Accuracy is validated against \oommf, and scalability shown by incorporation into a $1$-Mb $28$ $nm$ MRAM macro. The framework code is available upon request.

\bibliography{smacd_2021}
\bibliographystyle{noUrlIEEEtran}

\end{document}